\documentstyle[12pt]{article}

\def\rdots{\mathinner{\mkern1mu\raise1pt\vbox{\kern1pt\hbox{.}}\mkern2mu
   \raise4pt\hbox{.}\mkern2mu\raise7pt\hbox{.}\mkern1mu}}
\newcommand{\be}{\begin{equation}}
\newcommand{\ee}{\end{equation}}
\newcommand{\Z}{{\rm Z\kern-.35em Z}}
\newcommand{\bP}{{\rm I\kern-.15em P}}
\newcommand{\Q}{\kern.3em\rule{.07em}{.65em}\kern-.3em{\rm Q}}
\newcommand{\R}{{\rm I\kern-.15em R}}
\newcommand{\h}{{\rm I\kern-.15em H}}
\newcommand{\C}{\kern.3em\rule{.07em}{.65em}\kern-.3em{\rm C}}
\newcommand{\T}{{\rm T\kern-.35em T}}
\newcommand{\D}{{\kern-.5em /}}

\begin{document}

\openup 1.5\jot

\centerline{A Speculative Approach to Quantum Gravity$^*$}

\vspace{1in}
\centerline{Paul Federbush}
\centerline{Department of Mathematics}
\centerline{University of Michigan}
\centerline{Ann Arbor, MI 48109-1109}
\centerline{(pfed@math.lsa.umich.edu)}

\vspace{1in}

\centerline{Abstract}

	The bare bones of a theory of quantum gravity are exposed.  It may have the potential to solve the cosmological constant problem.  Less certain is its behavior in the Newtonian limit.

\vspace{3in}

$^*$ Contributed to the Symposium in Honor of Eyvind H. Wichmann, University of California, Berkeley, June 1999.

\vfill\eject

Throughout this discussion we work in Euclidean four-dimensional space-time.  The motivating example is of a scalar field $\phi(x)$ for which we want to construct a field theory within which
\be     \phi(x) \ge 0 \ \ \ \ {\rm all} \ \ \ \ x .	\ee
This is achieved by letting
\be	\phi(x) = e^{\psi(x)}	\ee
with $\psi$ a local quantum field.  We let the action for the field $\psi$ be given as
\be	S = \alpha \int \psi \Delta^2 \psi .   \ee
Then if $\alpha$ is correctly selected we can ensure
\begin{eqnarray}
	\Big< \phi(x) \phi(y) \Big> &=& \Big< e^{\psi(x)} e^{\psi(y)} \Big> = e^{C(x,y)} \nonumber \\
&\sim& \frac{1}{(x-y)^2} 
\end{eqnarray}
Here $C(x,y)$ is the two point correlation function of the $\psi$ field, and falls off logarithmically.

Analogously we wish to construct a field theory for $g_{\mu \nu}(x)$ within which the field $g_{\mu \nu}(x)$ has the correct signature for all $x$, i.e. it is positive definite:
\be	\eta^\mu g_{\mu \nu}(x) \eta^\nu \ge 0 \ \ \ {\rm all} \ \ \ x,\eta \; .	\ee
This we achieve by writing
\be	g_{\mu \nu}(x) = \Big( e^{A(x)}\Big)_{\mu \nu}	\ee
with $A$ a real symmetric matrix.  (Equations (5) and (6) parallel (1) and (2).)  We are led to expect that if the action of $A_{\mu \nu}$ is quadratic with four derivatives, the two point function of $g_{\mu \nu}$ may have a desired form.  With an action
 given as
\be	S_g = \int \sqrt{g} \Big( \alpha R_{\mu \nu} R^{\mu \nu} + \beta R^2 \Big)	\ee
the quadratic part of $S_g$, in terms of the $A_{\mu \nu}$ fields, is of this form, containing only four derivative terms.  The gauge fixing terms may also be chosen consistent with this requirement [1].  We then may hope that for a suitable choice of $\a
lpha$ and $\beta$ the two point function (as computed using only the quadratic terms of the action) may be of appropriate form.  It would be ideal if the values of $\alpha$ and $\beta$ of the full theory arose as an infrared fixed point of the renormaliza
tion group.

We make a number of comments and observations:
\begin{itemize}
\item[(1)] The theory is power-counting \underline{renormalizable}.  It is possible that a complete theory of renormalizability may be carried out as in [1].
\item[(2)] The theory formally solves the \underline{cosmological constant} problem, since it is invariant under $g_{\mu \nu} \rightarrow C g_{\mu \nu}$.  We believe this formal invariance can be extended through the quantization procedure.
\item[(3)] The theory is a \underline{large coupling} theory.  This is because $\alpha$ and $\beta$ chosen to pattern the two point function, will not be small.  The expression we get for the two point function, as we have discussed it, will be non-pertur
bative in a large coupling theory.  This will make calculations difficult and not necessarily reliable.  Perhaps we are dealing with the ``correct" theory, but in not the best formalism.
\item[(4)] \underline{Unitarity} is as elusive as usual in higher derivative theories.  It is perhaps present in low orders.  {\it We do not know how important unitarity is to a theory of gravitation}.
\item[(5)]  The heart of our present considerations is the \underline{gaussian approximation} to the two point function:
\be	\Big< e^{A(x)} e^{A(y)}\Big> \sim \int dA \ e^{- \frac 1 2 \int AC^{-1}A} e^{A(x)} e^{A(y)} \ee
written in a slightly schematic notation.  The integrals are very non-trivial because of the matrix nature of exponentials:
\be	\left( e^{A(x)}\right)_{ij} \ .	\ee
The evaluation of (8) is made possible (though still complicated) using the fourier transformation of the exponential in eq. (9) in terms of the entries of $A$:

\vfill\eject
\be
e^A = \int d\Omega e^{{\rm Tr}(AW)} \Bigg[ I + \frac {52}{3}( W - \frac 1 4 I)  - \frac{8}{3}   \ {\rm Tr}(AW)I + \frac{68}{3}  \ {\rm Tr}(AW) W-A
\ee
\[
+ \frac 1 6 \left[ - 4\ {\rm Tr}(A^2) + 46 ( {\rm Tr}(AW))^2 + 2A\ {\rm Tr}(AW)\right] W
\]
\[
- \frac 1 6 \left[ - \frac 1 2 \ {\rm Tr}(A^2)I + 3( {\rm Tr}(AW))^2 I + A^2 + 2A \ {\rm Tr}(AW) \right]
\]
\[
+ \frac 2 3 ( {\rm Tr}(AW))^3 W - \frac {1}{18} \ {\rm Tr}(A^3)W - \frac 1 6 \ {\rm Tr}(A^2) \ {\rm Tr}(AW)W \Bigg].
\]
This expression is from [2], and is written here for a traceless $4 \times 4$ matrix.  The integral $\int d\Omega$ is the integral over the unit sphere in $\C^4$, of a vector $v_i$ with respect to the normalized unitary-invariant measure.  And $W$ is the 
hermitian rank one projection given as
\be 	W_{ij} = v_i \bar v_j \ .	\ee
The program we envision is to select $\alpha$ and $\beta$ so that the gaussian approximation, eq. (8), yields a ``nearly reasonable" two point function; and to seek the ultimate two point function, via perturbative corrections.  The correct two point func
tions would  yield most of the tested properties of general relativity.  Remembering that we are working with a large coupling theory, we expect it to be difficult to get accurate predictions, to prove or disprove the theory.  This is a property we share 
with string theory.
\item[(6)]  We do not know what \underline{gauge condition} will prove best to study the present theory.  We have considered the condition
\be	\sum_\mu \partial_\mu A_{\mu \nu} = 0	\ee
as being one possibility.  For this gauge condition we have found a generalized BRS invariance [3], (more general than the generalized BRS transformations studied in [4]).  Of course for the usual harmonic gauge condition, the same BRS transformation as u
sed in [1] is expected to apply to the current model.
\item[(7)]  There has been some research on the \underline{cosmological} implications of actions as in eq. (7), [5].  For us the pressure is to find consistency with the Newtonian limit and the tests of general relativity.
\item[(8)] As a final subjective point we find many aspects of the present theory (so far studied only fragmentally) to be \underline{aesthetic}.
\end{itemize}

\bigskip
\bigskip

\centerline{REFERENCES}

\begin{itemize}
\item[[1]] K.S. Stelle, ``Renormalization of higher-derivative quantum gravity", {\it Phys. Rev. D}, {\bf 16}  953-969 (1977)
\item[[2]] P. Federbush, ``e to the A, in a New Way",  math-ph/9903006.
\item[[3]] P. Federbush, ``Some Generalized BRS Transformations", in preparation. 
\item[[4]] Satish, D. Joglekar, A. Misra, ``Relating Green's Functions in Axial and Lorentz Gauges Using Finite Field-Dependent BRS Transformations", hep-th/9812101.
\item[[5]] P.D. Mannheim, ``Conformal Gravity and a Naturally Small Cosmological Constant", astro-ph/9901219.
\item[  ]P.D. Mannhein, ``Curvature and Cosmic Repulsion", astro-ph/9803135. 
 \end{itemize}

\end{document}